\documentclass[prb,twocolumn,showpacs,showkeys,amssymb]{revtex4}

\usepackage{graphicx}% Include figure files
\usepackage{dcolumn}% Align table columns on decimal point
\usepackage{bm}% bold math

\begin{document}

%\preprint{}
%\tightenlines

\title{Discontinuities in the level density of small quantum dots under
 strong magnetic fields}
\author{Augusto Gonzalez}
\affiliation{Instituto de Cibernetica, Matematica y Fisica, Calle E
 309, Vedado, Ciudad Habana, Cuba}
\email{agonzale@cidet.icmf.inf.cu}
\author{Roberto Capote}
\altaffiliation[Permanent address: ]{Centro de Estudios Aplicados al
 Desarrollo Nuclear, AP 100, Ciudad Habana, Cuba}
\affiliation{Departamento de F\'{\i }sica At\'{o}mica, Molecular y Nuclear,
Universidad de Sevilla, Facultad de F\'{\i }sica, AP 1065, E-41080 Sevilla,
Spain}
\email{rcapote@us.es}

\begin{abstract}
\bigskip
Exact diagonalization studies of the level density in a six-electron quantum dot
under magnetic fields around 7 T (``filling factor'' around 1/2) are reported.
In any spin-polarization channel, two regimes are visible in the dot excitation
spectrum: one corresponding to interacting quasiparticles
(i.e. composite fermions) for excitation energies below 0.4 meV, and a second
one for energies above 0.4 meV, in which the level density
(exponentially) increases at the same rate as in the non-interacting composite-fermion
model.
\end{abstract}


\maketitle

The lowest energy states of relatively small quantum dots (number of electrons $N_e\ge 3$)
in strong magnetic fields have been qualitatively described whithin the
composite-fermion picture \cite{CF,Jain}. The position of cusps in the ground-state energy
as a function
of the angular momentum, and the spin quantum numbers of the low-lying excited states are nicely
reproduced by this theory. The residual interaction between composite fermions (CFs) is
expected to be a weak, contact interaction.

The present communication is aimed at giving a quantitative characterization
of the density of energy levels of small quantum dots in an energy interval below 1 -- 1.5
meV, where dozens or hundreds or states exist. Fully converged exact diagonalization results
for a 6-electron GaAs dot in magnetic fields corresponding to ``filling factors''
$\nu\approx 1/2$ are presented below. The error in computing energy eigenvalues is estimated
to be lower than 0.02 meV
for levels with excitation energies below 1 meV. The studied excitation energy range is small
as compared to the cyclotronic ($\hbar\omega_c=12$ meV), Coulomb (8.4 meV) or confinement
($\hbar\omega_0=3$ meV) energies of the model dot. Three Landau levels (LLs) are included in the
calculations, and a 25 meV cutoff in the energy of the non-interacting many-electron states
used as basis functions allows us to deal with hamiltonian matrices of dimension less than
850000, which are diagonalized by means of a Lanczos algorithm. The main result of the
paper is the existence of two regimes in the excitation spectra, corresponding to low
($\Delta E<0.4$ meV) and intermediate ($\Delta E>0.4$ meV) excitation energies, in
which the level density increases at different rates. Thus, at an
energy interval $\delta E$ around 0.4 meV the level density experiences a
``discontinuity''.

The model parameters are similar to those used in Ref. \onlinecite{physe}. The confinement
potential is parabolic. The bare Coulomb interaction is weakened by a factor 0.8 to
approximately account for quasi-bidimensionality (instead of exact bidimensionality).
The Zeeman energy is written in the form: 0.01432 $B~s_z$ meV ($B$ in Teslas and
$s_z=\pm 1/2$), corresponding to a 8 nm-width well in magnetic fields around 7 T \cite{boris}.

\begin{figure}[ht]
\begin{center}
\includegraphics[width=.9\linewidth,angle=0]{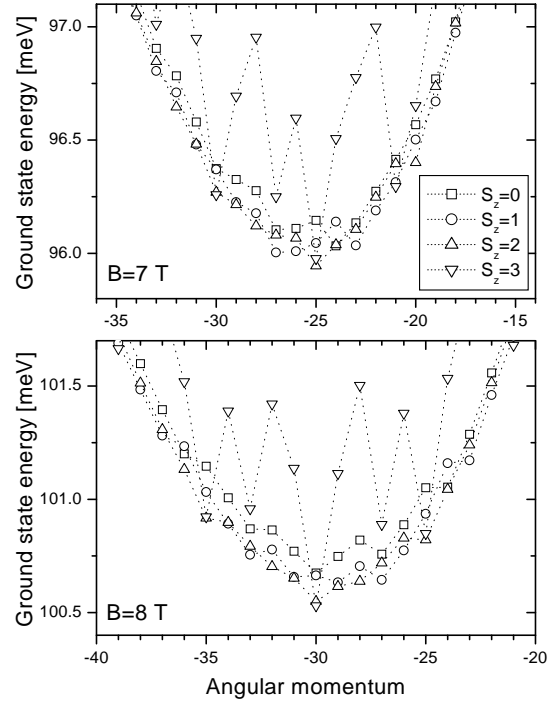}
\caption{\label{fig1}The lowest energy levels in each angular momentum
 and spin polarization sectors. Lines are guides to the eyes.}
\end{center}
\end{figure}

The lowest energy states in each angular momentum and spin polarization tower (the yrast
spectrum) for magnetic fields $B=7$ and 8 T are shown in Fig. \ref{fig1}. Energy jumps
between adjacent angular momentum states are about 0.6 meV in the spin-polarized case,
but roughly three times smaller in any other spin polarization sector. At this point it is
important to stress the role of the higher LLs in the energy eigenvalues. The absolute
contribution of the second and third LLs is around -0.4 meV, a magnitude
much greater than the Zeeman splitting (0.1 meV), and than the characteristic energy
spacing between states near the absolute minimum. Excitation energies are
pushed down 0.1 - 0.3 meV by the higher LLs.

\begin{figure}[ht]
\includegraphics[width=0.9\linewidth,angle=0]{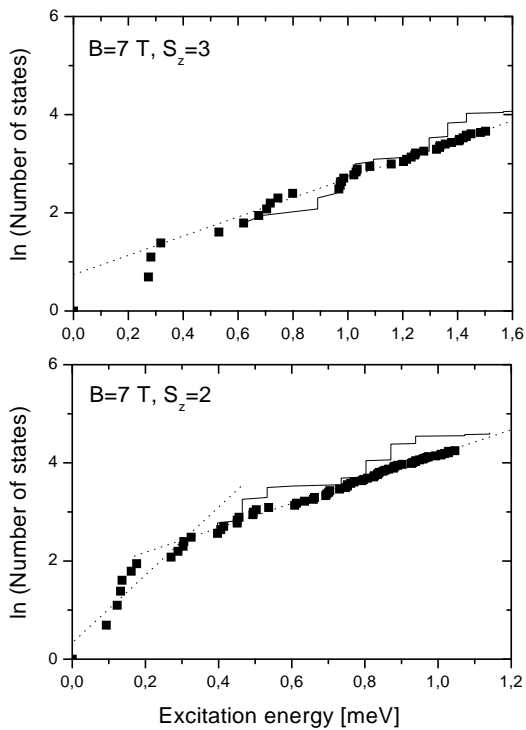}
\caption{\label{fig2}The logarithm of the number of levels as
 a function of the excitation energy at $B=7$ T: exact results
 (squares), constant temperature fits (dotted lines), and the shifted NICF
 curves (solid lines).}
\end{figure}

We show in Fig. \ref{fig2} the number of states as a function of the excitation energy,
$\Delta E$. In this figure, the reference energy is the minimal energy state whithin each
polarization sector. At low and intermediate excitation energies the curves are
well fitted by exponential functions \cite{physe},

\begin{equation}
n(\Delta E) = n_0 \exp \frac{\Delta E}{\Theta},
\label{ctaeq}
\end{equation}

\noindent
i. e. described by the so called ``constant temperature approximation''\cite{cta}.
In the polarized case, there are only a few states for low excitation energies,
thus we do not fit them. The fitting parameters $n_0$ and $\Theta$
are given in Table \ref{tab1}. Primed quantities correspond to the low-energy sectors.
Notice that the temperature parameter, $\Theta$,
changes very little on going from $S_z=2$ to $S_z=0$ for a given $B$. Notice also the
apparent jump of the temperature parameter at $\delta E$, which means a
discontinuity in the level density.

\begin{table}[ht]
\begin{tabular}{|c|c|c|c|}
\hline
$S_z$ &            & $B=7$ T & $B=8$ T \\
\hline
      & $\delta E$ & 0.30    & 0.45    \\
3     & $n_0$      & 2.099   & 1.712   \\
      & $\Theta$   & 0.510   & 0.443   \\
\hline
      & $n_0'$     & 1.398   & 1.614   \\
      & $\Theta'$  & 0.145   & 0.145   \\
2     & $\delta E$ & 0.35    & 0.30    \\
      & $n_0$      & 5.341   & 5.241   \\
      & $\Theta$   & 0.400   & 0.363   \\
\hline
      & $n_0'$     & 1.982   & 2.069   \\
      & $\Theta'$  & 0.107   & 0.085   \\
1     & $\delta E$ & 0.30    & 0.20    \\
      & $n_0$      & 11.57   & 10.99   \\
      & $\Theta$   & 0.398   & 0.344   \\
\hline
      & $n_0'$     & 1.506   & 1.635   \\
      & $\Theta'$  & 0.113   & 0.100   \\
0     & $\delta E$ & 0.40    & 0.25    \\
      & $n_0$      & 14.74   & 12.90   \\
      & $\Theta$   & 0.423   & 0.358   \\
\hline
\end{tabular}
\caption{The level density parameters $n_0$ and $\Theta$. Primed quantities
 correspond to excitation energies below $\delta E$.}
\label{tab1}
\end{table}

The non-interacting CF (NICF) understanding of the excitation spectrum starts
from a simplified 1LL picture in which the excitation energy is written in the form:

\begin{equation}
\Delta E=\hbar\omega_{_{CF}}~\Delta n_{_{LL}}+\hbar~\omega_0^2/\omega_c~\Delta L.
\end{equation}
\noindent
$\Delta n_{_{LL}}$ is the variation, with respect to the quasiparticle ground state,
of the effective LL occupation numbers.
\noindent
By definition, $\Delta L=|L|-|L_{gs}|$, where $L_{gs}$ is the ground-state angular
momentum. $\hbar\omega_c$ is the electron cyclotronic energy in GaAs, and
the effective cyclotronic energy of CFs is extracted from the polarized yrast spectrum,
$S_z=3$, as:

\begin{equation}
\hbar\omega_{_{CF}}=E(L_3+1)+E(L_3-1)-2 E(L_3),
\end{equation}

\noindent
where $L_3$ is the angular momentum of the lowest polarized state.
It gives $\hbar\omega_{_{CF}}=1.15$ and 1.19 meV at $B=7$ and 8 T, respectively.
The quasiparticles are supossed to occupy states in effective Landau levels, which
are separated by $\hbar\omega_{_{CF}}$. The angular momentum of the quasiparticle system
is $L_{_{CF}}$ which, due to the special form of the variational CF wave function,
is related to $L$ by

\begin{equation}
L=-N_e (N_e-1)+L_{_{CF}}.
\end{equation}

The NICF model gives qualitatively correct answers to questions like the
position of cusps in the yrast spectrum, the nature of the first excited states, etc.
One may ask for other visible manifestations of the NICF in the spectrum, and
indeed, it gives the correct $\Theta$ parameter at intermediate
excitation energies, $\Delta E>\delta E$. In other words, the NICF curve growths
exponentially with the same $\Theta$, although it is shifted with respect to
the actual spectrum. This fact is illustrated in Fig. \ref{fig2} also, where
the NICF curve at certain level (the sixth in the upper figure, for example) is
forced to meet the actual curve. From this point on the two spectra have
the same slope in average.

\begin{figure}[h]
\includegraphics[width=0.85\linewidth,angle=0]{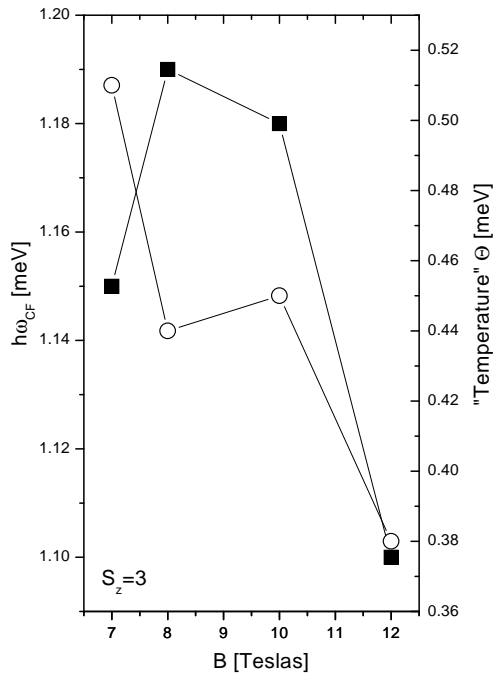}
\caption{\label{fig3} The magnitudes $\hbar\omega_{_{CF}}$ (squares) and $\Theta_{S_z=3}$
 (empty circles) as a function of magnetic field intensity $B$.}
\end{figure}

There is a natural interpretation of this behaviour. In the ground and
first excited states (low $\Delta E$),  the quasiparticles form compact clusters
\cite{Jain} and the interaction between CFs plays an important role. The
low-energy parameter $\Theta'$ is a consequence of this interaction. On the
other hand, at intermediate energies and assuming that the interaction is weak, one expects
the level density to be dominated by the relevant energy scales,
i. e. $\hbar\omega_{_{CF}}$ and $\hbar~\omega_0^2/\omega_c$. One may visualise this
regime in terms of Rydberg-like excitations, in which a few non-interacting CFs
orbit around a core of weakly interacting CFs. Energy differences will, of course,
follow the NICF rules.

The conclusion to be extracted from this figure is that traces of the effective
LL structure of CFs may also be looked for at intermediate excitation energies,
$\Delta E > 0.4$ meV. For $\Delta E < 0.4$ meV the transition to a regime of
weakly interacting quasiparticles takes place. Temperature parameters are
different from both sides of $\delta E$, thus a discontinuity in the level
density is expected.

The level density may be directly measured in Raman scattering
experiments under extreme resonance, where the Raman amplitude depends on
the density of energy levels in final states \cite{Raman}. Magnetoconductance
measurements under equilibrium \cite{equi} or non-equilibrium conditions
\cite{nonequi}, could also give evidence about the level density behaviour
at intermediate excitation energies.

Finally, we show in Fig. \ref{fig3} the magnitudes $\hbar\omega_{_{CF}}$ and $\Theta_{S_z=3}$
in a wider magnetic field intensity range, $7 \le B \le 12$ T. Near $B=12$ T, where the
``filling factor'' is aroud 1/3, there is a decrease of $\hbar\omega_{_{CF}}$ which
corresponds to an abrupt increase of the quasiparticle mass \cite{CFmass}. A similar
behaviour is observed in $\Theta$, showing that the level density parameter $\Theta$
depends on a negative power of $m_{_{CF}}$.

\begin{acknowledgments}
Part of this work was carried out at the Abdus Salam ICTP. A. G.
acknowledges the ICTP Associate and Federation Schemes for
support. R. C. acknowledges support from the
Ministerio de Educaci\'{o}n, Deportes y Cultura de Espa\~{n}a,
Secretar\'{\i}a de Estado de Educaci\'{o}n y Universidades.
\end{acknowledgments}

\end{document}